\documentclass[sigconf,natbib=true]{acmart}

\usepackage{paralist}
\usepackage{todonotes}
\usepackage[caption=false]{subfig}
\usepackage{multirow}
\usepackage[]{mdframed}

\AtBeginDocument{%
  }

\setcopyright{acmlicensed}
\copyrightyear{2018}
\acmYear{2018}
\acmDOI{XXXXXXX.XXXXXXX}
\acmConference[Conference acronym 'XX]{Make sure to enter the correct
  conference title from your rights confirmation email}{June 03--05,
  2018}{Woodstock, NY}
\acmISBN{978-1-4503-XXXX-X/2018/06}

\begin{document}







\title{REANIMATOR: Reanimate Retrieval Test Collections with  Extracted and Synthetic Resources}




\author{Björn Engelmann}
\affiliation{%
  \institution{TH Köln - University of Applied Sciences}
  \city{Cologne}
  \country{Germany}
}

\author{Fabian Haak}
\affiliation{%
  \institution{TH Köln - University of Applied Sciences}
  \city{Cologne}
  \country{Germany} 
}

\author{Philipp Schaer}
\affiliation{%
  \institution{TH Köln - University of Applied Sciences}
  \city{Cologne}
  \country{Germany}
}

\author{Mani Erfanian Abdoust}
\affiliation{%
  \institution{Science Media Center}
  \city{Cologne}
  \country{Germany}
}
\author{Linus Netze}
\affiliation{%
  \institution{Science Media Center}
  \city{Cologne}
  \country{Germany}
}

\author{Meik Bittkowski}
\affiliation{%
  \institution{Science Media Center}
  \city{Cologne}
  \country{Germany}
}
%

%

\renewcommand{\shortauthors}{Engelmann et al.}

\begin{abstract}

Retrieval test collections are essential for evaluating information retrieval systems, yet they often lack generalizability across tasks.
%
To overcome this limitation, we introduce REANIMATOR, a versatile framework designed to enable the repurposing of existing test collections by enriching them with extracted and synthetic resources.
%
REANIMATOR enhances test collections from PDF files by parsing full texts and machine-readable tables, as well as related contextual information.
It then employs state-of-the-art large language models to produce synthetic relevance labels. 
Including an optional human-in-the-loop step can help validate the resources that have been extracted and generated.
%
We demonstrate its potential with a revitalized version of the TREC-COVID test collection, showcasing the development of a retrieval-augmented generation system and evaluating the impact of tables on retrieval-augmented generation. 
%
REANIMATOR enables the reuse of test collections for new applications, lowering costs and broadening the utility of legacy resources.
\end{abstract}


\received{14 February 2025}

\maketitle

\section{Introduction}


Test collections have long been a cornerstone of information retrieval (IR) evaluation, forming the foundation of the TREC/Cranfield paradigm. These collections provide a controlled environment where retrieval systems can be evaluated against different datasets, queries, and relevance judgments. The underlying goal of such test collections is to facilitate the comparison of different IR approaches and foster improvements in retrieval effectiveness. However, despite their central role in IR research, test collections are not without their challenges and limitations.

One pressing concern with test collections is the considerable cost associated with their creation. Developing a high-quality test collection requires extensive time and financial resources, as it involves many manual editorial steps~\cite{DBLP:journals/ftir/Sanderson10}. 
The process involves gathering a representative set of documents, crafting meaningful topics, and securing reliable relevance assessments, often through expert annotators or crowd-sourced judgments. These efforts are substantial investments, making test collections costly to produce and maintain. For multi-modal collections~\cite{DBLP:conf/sigir/OardSDHMWRFGMKS04}, costs can reach many hundreds of thousands of USD. 
Given the rapid evolution of information retrieval needs and extending use cases, the ability to efficiently update or expand existing test collections is crucial for sustaining a meaningful evaluation ecosystem.

Moreover, test collections have been criticized for being artificial. 
New use cases for IR have emerged in recent years, pushing the boundaries of traditional test collections. Beyond document retrieval, there is a growing need to evaluate retrieval systems for tables, figures, and other non-textual content. 
Most test collections are static and uni-modal, limiting their applicability in assessing modern retrieval scenarios~\cite{DBLP:conf/ictir/KellerBS24}. Additionally, IR test collections are mainly constructed with a specific evaluation task in mind, which can lead to narrow and one-dimensional test scenarios. 
Mixed use cases such as question answering (QA) and retrieval-augmented generation (RAG) stress the limitations of current collections. These applications require test collections that encompass a diverse range of data types and query formats, yet most existing datasets remain tailored for conventional document retrieval. As a result, the field is confronted with the challenge of adapting evaluation methodologies to accommodate these novel retrieval tasks.

Despite the emergence of alternative evaluation methods -- including A/B testing, living labs, and simulations -- test collections continue to be the predominant approach for assessing IR systems. The primary reason for their continued relevance is their accessibility and reusability. Once a test collection is created, multiple research groups can take advantage of it, allowing reproducibility and comparative analyses in different studies~\cite{DBLP:conf/sigir/Breuer0FMSSS20}. This characteristic makes test collections a valuable asset for advancing the field, even considering their inherent limitations.

A fundamental objective of scientific inquiry is to build upon prior knowledge and extend existing work. In the context of IR evaluation, this means making test collections more sustainable and adaptable. The FAIR data principles\footnote{\url{https://www.go-fair.org/fair-principles/} (last accessed 6 February 2025} focus on Findability, Accessibility, Interoperability, and Reusability and offer a framework for enhancing the utility of existing test collections. Applying FAIR principles to IR evaluation fosters transparency, collaboration, and the cumulative progression of IR as a field.
Still, little research has been conducted to learn more about the repurposing of existing collections. 
Given the high costs and practical limitations associated with test collection development, it is only natural to explore ways to recycle and extend existing datasets. Instead of treating test collections as static entities, researchers should consider methods for reanimating and enriching these resources. This can involve augmenting document corpora with additional data, generating synthetic queries to simulate evolving information needs, and leveraging machine learning techniques to enhance relevance assessments. By systematically expanding all components of a test collection -- documents, topics, and judgments -- it becomes possible to create more dynamic and versatile evaluation frameworks that better reflect contemporary retrieval challenges. Current research is trying to synthesize test collections \cite{DBLP:conf/sigir/RahmaniCY0C24}, but whether this really helps to compensate for the previously mentioned concerns with respect to the artificial nature of test collections is still up for discussion.

In this resource paper, we propose a framework for revitalizing existing retrieval test collections through a mixed-methods approach, where we include modern information extraction techniques and LLM-created synthetic resources. 
REANIMATOR is a toolkit developed for \textbf{RE}vitalizing \textbf{AN}d \textbf{IM}proving \textbf{A} \textbf{T}est c\textbf{O}llection for \textbf{R}etrieval.
This includes parsing of full texts and tables, preparation for retrieval tasks, and automated synthetic relevance assessment.
Our methodology leverages existing test collections while introducing mechanisms to extend their scope and adaptability. By integrating extracted resources from real-world corpora and generating synthetic elements to complement missing components, we aim to create test collections that remain relevant in the face of evolving retrieval paradigms. Through this effort, we seek to bridge the gap between traditional test collections and the demands of modern IR applications, ensuring that evaluation methodologies continue to support innovation in the field.




Our research contributions are
\begin{inparaenum}[(a)]
\item a novel framework for automatically enriching a literature collection with machine-readable parsed tables, captions, and in-text references, as well as automatic generation of synthetic relevance labels for a wide range of retrieval tasks,
\item an application of REANIMATOR for reanimating the TREC-COVID test collection with tables, corresponding context, and relevance judgments\footnote{Source code, data, and resources, including utilized LLM prompts and the test collection are available at \url{https://github.com/irgroup/Reanimator}.},
\item an evaluation of the use of tables in RAG. 
\end{inparaenum}
To encourage reproducibility and facilitate further research, we will publicly release our implementation under the MIT license, along with all relevant data and resources in our GitHub repository. This includes document chunks for retrieval, poolings, relevance assessments, retrieved and generated (RAG) answers, and the retrieval models used in our experiments.

\section{Related Work}



\textbf{Recycling test collections} is essential for a FAIR data principle-driven IR ecosystem.
\citet{scells_reduce_2022} propose a ``Green Information Retrieval'' framework that emphasizes conserving research resources by reducing, reusing, and recycling existing software artifacts.
Within this paradigm, \emph{reuse} refers to deploying data, code, or models for essentially the same task they were designed for, whereas \emph{recycle} involves repurposing these artifacts for a new task or context with minimal modifications.
In our context, these principles directly inform our effort to ``reanimate'' existing test collections.

Ensuring that IR datasets are easy to locate, access, integrate, and repurpose is a requirement for the IR community to maximize the long-term impact of test collections and reduce the burden of developing new ones from scratch. Platforms like the TREC Browser~\cite{DBLP:conf/sigir/0002VS24} or \texttt{ir\_metadata}~\cite{DBLP:conf/sigir/MacAvaneyYFDCG21} help to locate and access existing collections and corresponding work. Instead of reinventing the wheel, the TREC Browser can help identify relevant test collections. 

Reusing test collections involves adding new topics and relevance assessments or adapting them for different application domains, such as transitioning from information retrieval evaluation to recommender systems. These methods typically depend on manual effort. An alternative approach to constructing test collections was introduced in the Social Book Search track of CLEF \cite{DBLP:conf/clef/KoolenKPD13}. This approach leveraged the INEX Amazon/LibraryThing collection, enriching it with external content from the LibraryThing website to derive topics and relevance assessments. This method, which involves developing specialized web crawlers, represents a more technical approach to obtaining additional data. 
Another method is to use internal content already available in the (original) document collection. With the help of information extraction methods, it was shown how additional metadata could be leveraged from the original document collection and how this would enable new evaluation scenarios in the domain of academic search~\cite{DBLP:conf/clef/SchaerN17,DBLP:conf/ecir/LarsenL16,DBLP:journals/scientometrics/MutschkeMSS11}.
Another example of a recycling test collection was the TREC 2017 Common Core built on existing materials. Topics from TREC Robust04 were re-used and transferred to a new document collection. The (re-)construction approach involved modeling the relevance assessment process as a multi-armed bandit problem~\cite{DBLP:conf/trec/AllanHKLGV17,DBLP:conf/ceri/LosadaPB18}. This method aims to identify relevant documents with minimal effort, thereby reducing the overall cost of building test collections. 

\textbf{Tables} are underrepresented elements in IR test collections, although they are useful for cases like search~\cite{engelmann_simulating_2023,shraga_web_2020,wang_context_2015}, question answering~\cite{herzig_open_2021} or fact verification~\cite{chen_tabfact_2019}.
In academic research, study results, findings, and methodological approaches are often presented concisely as tables in scientific literature.
Despite this, academic retrieval systems often overlook table information content, headers, and table context information, like captions and in-text references, as tables typically aren't included in standard test collections.
Existing collections like the PMC Gold standard table corpus~\cite{habibi_tabsim_2020} or TableArXiv~\cite{gao_scientific_2017} do not include context information such as captions and in-text references and employ manual labeling.
To the best of our knowledge, our framework is the first to allow for the systematic construction of table retrieval test collections from PDF documents that include context information and assign synthetic relevance labels.

\textbf{Synthetic relevance assessments} are a complement or substitution for human relevance assessments, which are known to be time-consuming and expensive.
With the recent advances in language models, synthetic relevance labels assigned by LLMs have become a viable option.
Despite concerns~\cite{soboroff2024dontusellmsmake}, there are recent studies that argue that by utilizing a suitable setup and prompting, as well as a human-in-the-loop approach for validation, human-level assessments can be achieved~\cite{thomas2024largelanguagemodelsaccurately, umbrela}.

\section{REANIMATOR Framework}

\begin{figure}[t]
    \centering
    \includegraphics[width=1\linewidth]{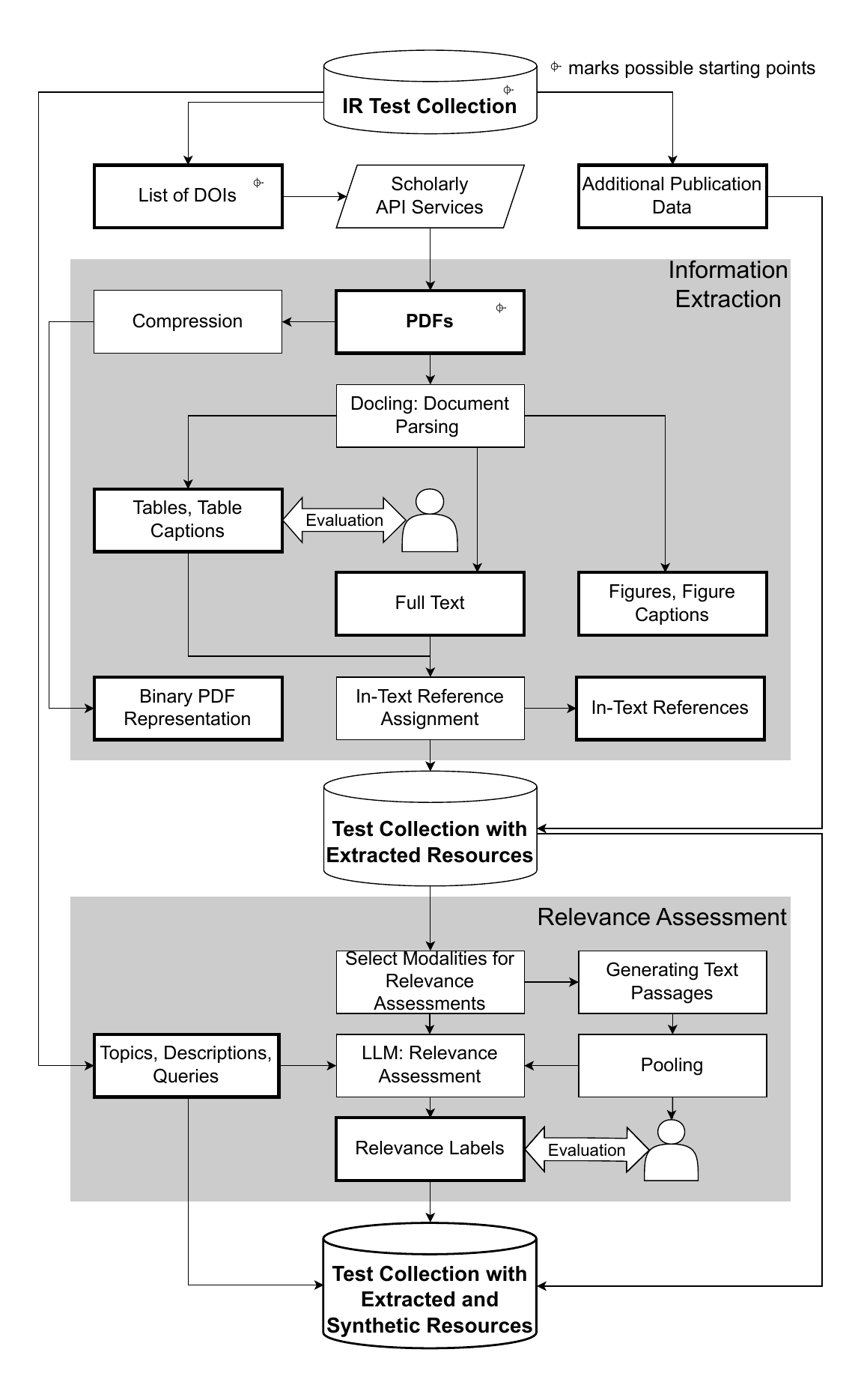}
    \caption{Outline of the methodology employed in the REANIMATOR framework for enriching existing retrieval test collections and constructing new test collections from a set of PDF files.}
    \label{fig:framework}
\end{figure}

This section introduces REANIMATOR, our framework for revitalizing and improving test collections by extracting and synthesizing resources, as outlined in \autoref{fig:framework}.
Starting from either an existing test collection, a list of DOIs, or a set of unaltered documents (in PDF, HTML, or other data formats supported by our information extraction framework), REANIMATOR consists of two primary components: 
\begin{inparaenum}[(a)]
    \item extraction of document content and
    \item automated relevance assessment.
\end{inparaenum}

\subsection{Setup and Input Data}\label{sec:reanimator-setup}
The default use case of REANIMATOR starts with an existing retrieval test collection, making it suitable for a new retrieval task.
We recommend using \texttt{ir\_datasets}~\cite{DBLP:conf/sigir/MacAvaneyYFDCG21} for their ease of use and uniform data format.
Any current test collection is usable, provided there's a list of document sources (like DOIs or URLs) or a set of supported document file types with a clear identifier linking to the collection data. 
Examples for this kind of collection come from the domain of academic search like TREC-COVID\footnote{\url{https://ir.nist.gov/trec-covid/index.html}}, iSearch\footnote{\url{https://sites.google.com/view/isearch-testcollection/}}, or collections used in the
TREC Biomedical Tracks\footnote{\url{https://www.trec-cds.org/}} (like the Precision Medicine or Clinical Decision Support Tracks).  
When starting with an existing test collection or a list of DOIs, if no PDF files are provided, REANIMATOR collects PDF URLs through various scholarly API services (OpenAlex\footnote{\url{https://docs.openalex.org/}}, Wiley API\footnote{\url{https://onlinelibrary.wiley.com/library-info/resources/text-and-datamining}}, and Unpaywall\footnote{\url{https://unpaywall.org/products/api}}). 
While this approach does not retrieve documents behind paywalls, it does allow access to most publicly available documents.
Full texts and any additional metadata, such as titles and author information, are also adopted from the test collection.
Available topics and related information can be reused if suitable for the target application.
In principle, REANIMATOR can construct a new test collection from a set of full-text documents. 
In that case, providing metadata is optional. 
Topics, as well as optional descriptions and narratives, must be specified separately. Currently, this has to remain as work in the future as we focus on updating available collections. 

\subsection{Information Extraction}
\label{sec:information_extr}
Extracting information from PDF documents forms the backbone of REANIMATOR. 
Different retrieval tasks require a variety of document-related resources, so REANIMATOR is designed to accommodate many aspects of documents in different forms.
For parsing the full-text documents, we utilize Docling~\cite{Docling}, which is able to parse PDF, DOCX, XLSX, HTML, or many other document formats. 
This information extraction framework includes an advanced PDF understanding incl. page layout, reading order, table structure, code, formulas, or image classification.

Especially for legacy retrieval test collections, full texts are often absent or only partially included. 
Often, if full texts are part of the test collection, the quality of the texts extracted in the past cannot match the quality of current parsing implementations.
Hence, full-text extraction is an important component of REANIMATOR, laying the foundation for tasks such as passage retrieval or RAG.

REANIMATOR also leverages Docling to extract resources of various modalities. 
We refer to the term modalities as the different data formats in which information can appear, such as texts, tables, figures, and other structured or unstructured representations. 
By utilizing Docling, REANIMATOR ensures seamless extraction and integration of relevant data across these diverse modalities.

The content and structure of tables are parsed, and captions are recognized and linked to their respective tables. 
Because the expanded context of tables can be relevant for tasks like table retrieval, REANIMATOR locates in-text references to tables. 
By extracting the table name from its caption and identifying mentions in the full text, we can add segments of the text that potentially provide additional meta-information about the data presented or any analyses and summaries not included in the caption.
REANIMATOR offers a human-in-the-loop option to verify the quality of extracted tables and their corresponding context.
In addition to tables, figures and their captions are extracted.
All documents, tables, figures, and extracted information are stored in a relational database to enable controlled and isolated access.
PDFs are compressed into binary representations for quick access, for example, when rendering figures in various resolutions.

\subsection{Relevance Assessment}\label{sec:relevance_ass}
The resulting (extended) test collection can be passed to the REANIMATOR relevance assessment module. 
If only the extracted resources are needed, for example, full texts for an existing test collection, relevance assessment can be skipped.
However, if relevance labels are needed for the extracted resources, REANIMATOR assigns synthetic relevance labels.
This step is necessary since we cannot assume that the extracted tables, figures, etc., share the same relevance labels as the original document from which they were extracted.
While we can assume that a document is still relevant when it is extended by additional metadata, the same is not true for the subsequent document parts that belong to it.
In addition to that, parts of a document can be highly relevant for a topic, even if the document as a whole is not relevant or only partially relevant, which applies particularly for longer documents.  
Therefore, synthetic relevance labels are assigned to instances of the user-selected modality by an ensemble of interchangeable open and closed LLMs, ensuring the collection's suitability for a wide range of retrieval tasks.
Tables, table context information, and text passages can be selected as modalities to be judged.
Chunking is essential for RAG and passage retrieval, and REANIMATOR offers various configuration options to accommodate different requirements. 
Figures and entire full texts can also be included by choosing visual models or models with a larger context length, respectively.
Our framework creates a pooling for each modality and topic that aggregates rankings from multiple retrieval models to generate a diverse candidate list for retrieval evaluations.
This process and an exemplary implementation are explained in more detail in Section~\ref{sec:rag}.

Relevance assessment is facilitated by UMBRELA~\cite{umbrela}, an open-source, state-of-the-art relevance assessing framework proven effective for TREC-style topics\footnote{\url{https://github.com/castorini/umbrela}}.
An UMBRELA prompt was used to categorize each candidate text passage according to four levels of relevance (see Figure~\ref{fig:umbrela-prompt}).

\begin{figure}
    \centering
    \begin{mdframed}[innerleftmargin=6pt,innerrightmargin=6pt]
\small\textit{\textbf{Irrelevant}: Passage has nothing to do with the query.\newline
\textbf{Related}: Passage seems related to the query but doesn't answer it.\newline
\textbf{Highly relevant}: Passage has some answer for the query, but the answer may be a bit unclear, or hidden amongst
extraneous information.\newline
\textbf{Perfectly relevant}: Passage is dedicated to the query and contains the exact answer.}
    \end{mdframed}
    \caption{Four levels of relevance used to formulate the original UMBRELA prompt for assessing the relevance of extracted passages and tables.}
    \label{fig:umbrela-prompt}
\end{figure}

REANIMATOR, by default, allows for a wide range of different models to be used for relevance assessment.
For an example, see Section ~\ref{sec:reanimating-trec-covid}, where we evaluate different models for assessment of text passages and table relevance assessment.
Depending on the chosen modalities, different types are more or less effective and efficient.
Relevance labels can be evaluated through a labeling tool provided with REANIMATOR, which allows the calculation of inter-annotator agreements between human and LLM-based relevance assessments.


\section{Reanimating TREC-COVID}\label{sec:reanimating-trec-covid}


To demonstrate the functionality of REANIMATOR, we process the TREC-COVID test collection to make it suitable for table retrieval and a table-considering RAG approach, creating TREC-COVID+.
TREC-COVID is based on the document collection CORD-19, which consists of around 193k scientific articles related to COVID-19~\cite{Wang2020Cord19}. These scientific articles include many tables that are not directly available or retrievable in the original collection. While the text content of tables is included in the precomputed SPECTER document embeddings for each CORD-19 paper and in the collection of JSON files that contain full-text parses of a subset of CORD-19 papers - the tables' specific context and structure are not available. Therefore, we took TREC-COVID as a case study to apply REANIMATOR and demonstrate the feasibility of our approach. Other collections that are of a non-academic nature, such as typical newswire collections (like the New York Times or Washington Post corpora), might also have been considered, but we expected them to include fewer tables per article in comparison to a set of scientific articles. 

The TREC-COVID collection includes approximately 136k unique DOIs, as some DOIs are associated with multiple versions of the same article. 
In our dataset construction, we included only one article per DOI.
Aside from the documents, TREC-COVID contains about 70k high-quality relevance scores for 50 detailed topics~\cite{Voorhees2020TrecCovid}.
However, these relevance labels are for documents and (available parts of) their full texts, neither passages nor tables. 
We chose TREC-COVID because it is particularly suitable for enrichment with tables and for RAG.
The biomedical domain contains a high diversity of academic research, and tables can contain highly relevant information in the context of COVID-19. 
Further, with the diverse and highly technical medical research represented in CORD-19, the corpus is a robust proving ground for assessing how well RAG models can locate and synthesize specialized information.
Applying REANIMATOR, we provide artifacts that expand the TREC-COVID test collection by adding extracted tables and corresponding context, text passages, and relevance labels to make it suitable for the intended purpose of table retrieval and RAG, which will be presented in more detail in Section~\ref{sec:rag}.


\subsection{Creating TREC-COVID+}\label{sec:TREC-COVID+}
Processing TREC-COVID with REANIMATOR yielded 64,358 publicly available PDF documents via the APIs described in Section ~\ref{sec:reanimator-setup}. This is nearly half of the available unique DOIs in TREC-COVID. From these PDFs, we extracted full texts, tables, table captions, and their in-text references for the use case of table retrieval. We also extracted text passages for the later RAG use case. Additionally, tables and passages were automatically labeled with REANIMATOR's LLM-based relevance assessment pipeline.

\paragraph{Table extraction} REANIMATOR was able to extract 144,206 tables, 99,286 table captions, and 77.252 in-text references. This represents a substantial dataset, although for 44,920 tables (31\%), captions are missing, and for 66,954 tables (53\%), in-text references could not be identified. The absence of captions presents a challenge for automated reference extraction, as in-text mentions rely on clearly labeled tables.
Using the parsing evaluation module of our toolkit, we investigated how well Docling in the current version v2.15.0 used for REANIMATOR parses tables from the PDF files (see Section~\ref{sec:quality-resources}) and how well our relevance assessment module can synthesize relevance judgments (see Section~\ref{sec:quality-synthetic}).

\paragraph{Passage Extraction}
Full texts must be chunked into text passages to use them in RAG. 
In line with the recommendations set by Wang et al.~\cite{wang_searching_2024}, we employ a chunk size of 512 characters with an overlap of 100 characters.
This results in a total of 8,475,683 passages for the parsed documents.


\paragraph{Pooling}
We build upon the approach of Moffat et al.~\cite{10.1145/2806416.2806606}, adopting their pooling suggestion with query variants to generate candidate tables and text passages based on different query variations per topic.
We use two retrieval models, BM25 and cosine similarity based on embeddings generated with the \textit{text-embedding-3-small} model\footnote{\url{https://platform.openai.com/docs/guides/embeddings/embedding-models}}. 
TREC-COVID comes with 50 topics, each consisting of a title, a description, and a narrative. 
Retrieval queries are formulated as a combination of title and description.
For both retrieval models, we generate five query variants with \textit{gpt-4o-2024-11-20}
\footnote{
Prompt used: \texttt{You are an AI assistant specialized in retrieving scientific information.  
Your task is to generate five distinct rephrasings of the user question so they can be  
effectively used with both sparse (e.g., BM25) and dense (e.g., cosine similarity) retrieval methods.  
Make sure each rephrasing captures different potential keywords, synonyms, or contexts  
specific to scientific research. Provide the five versions separated by newlines.}  
\newline
\texttt{Original question: \{question\}}
}, resulting in six rankings for each retrieval model, for each of the two modalities: tables and passages.
Using Reciprocal Rank Fusion~\cite{10.1145/1571941.1572114}, we compile a top-200 list for each of the 50 topics. This comprehensive ranking incorporates all 12 query variant rankings.
This results in two overall rankings for each topic and 20,000 relevance assessment pairs: top-200 ranking for 50 topics and two modalities.

\paragraph{Relevance Assessment} 
We label the pooled tables and text passages based on four levels of relevance (\emph{irrelevant, related, highly relevant,} and \emph{perfectly relevant}) in accordance with the UMBRELA-style prompting framework (as described in \autoref{sec:relevance_ass}).
We use the full TREC-COVID topic information, title, description, and narrative.
We employ a diverse set of open-source and proprietary large language models for relevance assessment.
Specifically, we include \textit{Qwen2.5-14B-Instruct}, \textit{Google\_gemma-2-9b-it}, \textit{Microsoft\_phi-4},\textit{Mistral-7B-Instruct-v0.3}, \textit{Mistral-Small-Instruct-2409}, and \textit{Falcon3-7B-Instruct}, that are run on a local machine.
Additionally, the closed-model variants \textit{o3-mini-2025-01-31}, \textit{gpt-4o-mini-2024-07-18} and \textit{gpt-4o-2024-11-20} are used.
Table prompts include the table caption and any relevant in-text references that clarify numerical content or methodological details.
This setup ensures that each model can assess table relevance with all necessary background information.
In addition to the relevance labels, we produced a majority vote label of all three GPT models.

\paragraph{Costs} The costs per relevance assessment are listed in Table~\ref{tab:inter-rater}. The total cost for the proprietary OpenAI models are 38\$ (28\$ for tables and 10\$ for passages) for \textit{gpt-4o}, 22\$ (15\$ + 7\$) for \textit{o3-mini}, and 2.27\$ (1.66\$ + 0.61\$) for \textit{gpt-4o-mini}. 
Experiments were conducted on a workstation running Ubuntu 20.04 LTS, powered by an AMD EPYC 7443P CPU (48 cores, 2.85 GHz) with 256 GB of memory. A single NVIDIA RTX A6000 GPU (48 GB memory) was used for all GPU-accelerated computations. 
The cost for the human annotators can only be roughly estimated. The typical assessment session for 125 tables and 125 passages was about 4 hours long. This would be an actual duration of roughly 1 minute per judgment (1.2 for tables and 0.8 for passages). 

\paragraph{Availablity} TREC-COVID+ is fully available online\footnote{\url{https://drive.google.com/drive/folders/1IqhijGWffGQ5ZjE7JrGTDAwPq_PGFVXD?usp=sharing}}, including all the extracted and generated resources, but without the original full texts. These can be crawled using the scripts and methods included in REANIMATOR.


\subsection{Quality of Extracted Resources}
\label{sec:quality-resources}
\begin{table}[t]
\caption{Evaluation of table parsing quality on a subset of the full table set. Misclassified tables are excluded.}
\small
\begin{tabular}{l c c c c c}
\toprule
        & perfect & good  & ok   & bad & total \\ 
        \midrule
count   & 287   & 115 & 30 & 24  & 456 \\
percent & 62.94\%   & 25.22\% & 6.58\% & 5.26\%  & 100\%\\
\bottomrule
\end{tabular}
\label{tab:parsing_quality}
\end{table}

\begin{table}[t]
\caption{Evaluation of table caption parsing quality on a subset of the full table set. Tables without captions and misclassified tables are excluded.}
\label{tab:caption_parsing}
\small
\begin{tabular}{l c c c c}
\toprule
        & perfect & not recognized & other & total \\ 
\midrule
count   & 323     & 76                     & 4     & 403   \\
percent & 80.15\%   & 18.86\%                  & 1\%     & 100\%\\
\bottomrule
\end{tabular}
\end{table}

We randomly sampled and manually labeled 500 tables from all parsed documents.
Of these, 44 were misclassifications (e.g., figures or parts of text interpreted as tables, often references), leaving 456 actual tables to be evaluated.
\autoref{tab:parsing_quality} provides a summary of the analysis of the quality of the parsing of valid tables.
Among the valid tables, 62.94\% were parsed perfectly, while 25.22\% were deemed ``good'', indicating only minor structure- or content parsing issues.
An additional 6.58\% were labeled as ``ok'', reflecting more noticeable but minor imperfections (e.g., missing rows, suboptimally merged multi-indices).
Only 5.26\% were classified as ``bad'', denoting significant parsing errors like missing or merged columns or mangled parsed structure.
Overall, these findings suggest that roughly 95\% of actual tables are at least substantially correct, highlighting the reliability of the parsing pipeline despite occasional misclassifications and inaccuracies.

Among the 456 valid tables, 53 were identified as having no caption, leaving 403 instances for caption analysis.
As shown in Table~\ref{tab:caption_parsing}, 80.15\% of these captions were extracted perfectly, while captions for 18.86\% of the tables were missed.
Only around 1\% of cases fell into the ``other'' category, which includes wrong caption assignment and incomplete parsing.
Overall, these results demonstrate that captions are reliably captured for the majority of tables.

\subsection{Quality of Synthetic Resources}
\label{sec:quality-synthetic}

We employ relevance assessments of eight annotators to evaluate the synthetic labels. 
Annotators are computer scientists of various experience levels.
For each topic, the top five and bottom five passages and tables are selected for human labeling.
By selecting both top- and bottom-ranked elements, we aim to achieve a more balanced distribution of relevant and non-relevant items.
Each annotator labels 125 passages and 125 tables. 
\autoref{tab:inter-rater} and \autoref{tab:inter-rater_binary} show the average Cohen's Kappa for human-human and human-LLM inter-rater agreements for 4-level relevance and binary assessments, respectively.  
For the binary labels, we introduce a ``surrogate label'' that uses the relevance label of the corresponding TREC-COVID document for the tables.
This is based on the assumption that tables from a relevant document are likely to be relevant, too.

For the 4-level relevance assessment, human inter-rater agreement scores for passages and tables are almost identical, with both at around 0.35. 
The best models perform on par with human raters, on similar Cohen's Kappa score levels.
For binary relevance assessment, the human inter-rater agreement is notably higher for passages. 
Better performing models show on par or higher Cohen's Kappa scores than average inter-human scores for binary relevance assessment.
Overall, given the stronger models, our results are on par with the original work of UMBRELA and their extensive evaluation scheme for synthetic relevance assessments across different TREC collections~\cite{umbrela}.

The surrogate labels come with no extra costs but are outperformed by the human assessments and by all but the worst-performing models. While the surrogate labels can be understood as a naive baseline, the experiments show the limitations of recycling old collections and their labeled data. Re-assessing the relevance of new artifacts should always be considered.

\begin{table}[t]
\caption{Cohen's Kappa inter-rater agreement of human raters and language model labels and cost per relevance assessment of LLMs for 4-level relevance assessment. Costs are measured in API  costs for proprietary models (in USD cents) or time for local/open source models (in seconds). Bold numbers denote the best values for each variant and column.}
\label{tab:inter-rater}
\small
\begin{tabular}{l c c c c}
\toprule
                            & \multicolumn{2}{c}{Cohen's $\kappa$} & \multicolumn{2}{c}{Cost/assessment}      \\
                            & Tables         & Passages       & {Tables}       & Passages          \\ 
\midrule
human                       & 0.355          & 0.351          &   70.2 s            &    45.6 s              \\ 
\midrule
gpt-4o-2024-11-20           & 0.376          & 0.289          & 0.280 \textcent          & 0.100 \textcent          \\
gpt-4o-mini-2024-07-18      & 0.394 & 0.337          & \textbf{0.017 \textcent} & \textbf{0.006 \textcent} \\
o3-mini-2025-01-31          & 0.384          & \textbf{0.342} & 0.150 \textcent          & 0.070 \textcent          \\
majority\_vote                   & \textbf{0.416}           & 0.333            &                         &          \\
\midrule
Falcon3-7B-Instruct         & 0.197          & \textbf{0.325} & \textbf{0.629 s} & \textbf{0.457 s} \\
google\_gemma-2-9b-it       & 0.308          & 0.280          & 0.966 s          & 0.715 s          \\
microsoft\_phi-4            & 0.191          & 0.218          & 2.773 s          & 4.740 s         \\
Mistral-Small-Instruct-2409 & 0.101          & 0.182          & 2.807 s          & 1.750 s          \\
Mistral-7B-Instruct-v0.3    & 0.291          & 0.155          & 3.053 s          & 2.741 s          \\
Qwen2.5-14B-Instruct        & \textbf{0.370} & 0.319          & 5.911 s          & 5.288 s          \\
\bottomrule
\end{tabular}
\end{table}

\begin{table}[t]
\caption{Cohen's Kappa inter-rater agreement of human raters and language model labels for binary relevance. Bold numbers denote the best values for each variant and column.}
\label{tab:inter-rater_binary}
\small
\begin{tabular}{l c c}
\toprule
  & \multicolumn{2}{c}{Cohen's $\kappa$}\\
                                     & table   & passage     \\
\midrule
human                              & 0.491    & 0.567        \\
surrogate                          & 0.355  & 0.237 \\
\midrule
gpt-4o-mini-2024-07-18            & 0.576     & \textbf{0.570}  \\
gpt-4o-2024-11-20                  & 0.513    & 0.551  \\
o3-mini-2025-01-31                 & 0.556    & 0.514  \\
majority\_vote                    & \textbf{0.584}     & 0.558  \\
\midrule
google\_gemma-2-9b-it             & 0.414      & \textbf{0.534} \\
Qwen2.5-14B-Instruct        & \textbf{0.537}     & 0.498  \\
Falcon3-7B-Instruct        & 0.380     & 0.481 \\
Mistral-Small-Instruct-2409 & 0.188 & 0.422  \\
microsoft\_phi-4                  & 0.272     & 0.408  \\
Mistral-7B-Instruct-v0.3 & 0.458    & 0.142 \\
\bottomrule
\end{tabular}
\end{table}

\section{A Table Retrieval and RAG Case Study for TREC-COVID+}\label{sec:rag}

\begin{figure}
    \centering
    \includegraphics[width=1\linewidth]{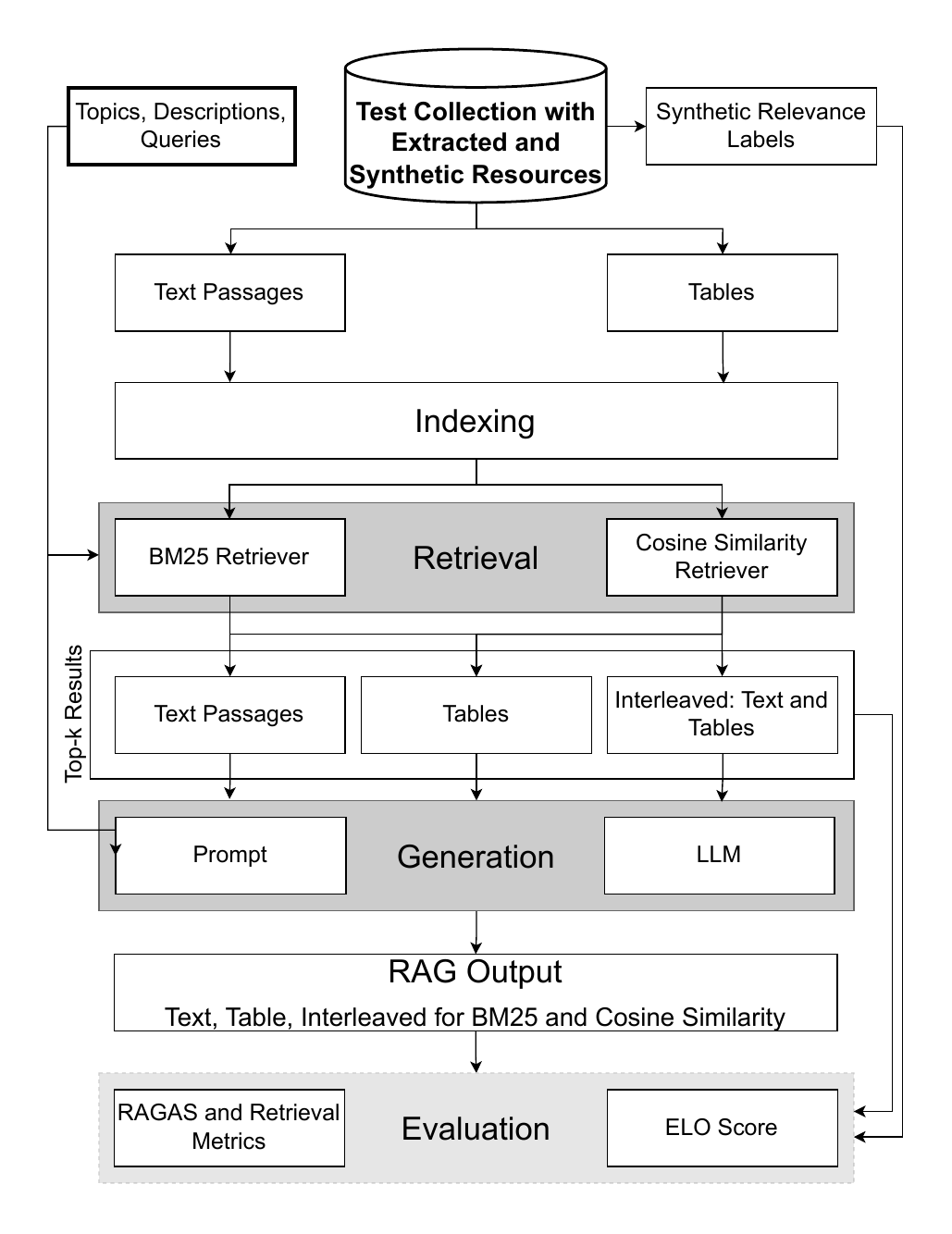}
    \caption{RAG experimental setup for a REANIMATOR-generated test collection based on TREC-COVID.}
    \label{fig:rag_flow}
\end{figure}

As an exemplary application of REANIMATOR, we use the reanimated TREC-COVID+ collection with extended table resources in \begin{inparaenum}[(a)]
\item a text/table retrieval and 
\item a RAG case study.
\end{inparaenum}
We perform RAG experiments without requiring human effort \cite{es_ragas_2023}. 
The general setup involves a pipeline where the system processes a query or question, retrieves relevant text passages and tables, and then combines these with the original query as context. 
This enriched context is then used by an LLM to generate the final answer \cite{es_ragas_2023, xiong_benchmarking_2024, yu_evaluation_2024}.
\autoref{fig:rag_flow} outlines the methodology of our RAG case study.
Given the numerous possible configurations and parameters in both retrieval and RAG experiments, we adopted default settings and best practices from the literature \cite{yu_evaluation_2024, gao_retrieval-augmented_2024, wang_searching_2024, gu2025surveyllmasajudge}—not to achieve optimal performance, but to establish a baseline that enables reproducible experiments.

\subsection{Retrieval Setup}\label{sec:retrieval_setup}
To investigate the impact of tabular data on RAG, we compare three distinct retrieval modality configurations: \begin{inparaenum}
    \item text-only,
    \item table-only, and
    \item interleaved retrieval, encompassing both text and tables.
\end{inparaenum} 
In each configuration, the top-10 ranked elements are retrieved from two separate indices, each built using a different retrieval model—BM25 or cosine similarity.
The rankings and the query are given to an LLM to generate the final answer. 
This setup enables a systematic comparison of tabular content alone with purely textual segments and whether interweaving text with tables yields any additional benefits.

\subsection{RAG Setup}\label{sec:rag_setup}
%
In the augmented generation phase of our RAG pipeline, we employ \textit{claude-3-5-Sonnet-20241022} as the language model responsible for synthesizing responses based on the retrieved context and input query.
We selected an LLM model that does not belong to the GPT family because in the ``LLM as a Judge'' paradigm, greater diversity in the generation-to-judge relationship is recommended. Gu et al. describe this problematic phenomenon as self-enhancement bias \cite{gu2025surveyllmasajudge}.
For each of the 50 TREC-COVID topics, we generate answers using the top-10 retrieved results from both retrieval models (BM25 and Cosine Similarity) across three different modality configurations.
This results in a total of $50 \times 2 \times 3 = 300$ retrieval-based input sets.
To account for potential variations in model outputs, we generate five independent RAG responses per combination, yielding a total of 1,500 generated answers.

Each prompt consists of the original query title and description along with the corresponding top-10 ranked context retrieved for the specific modality configuration and retrieval approach. For our experiments, we employed a structured prompt to generate answers based on the provided documents\footnote{Prompt used: \texttt{You are a helpful AI assistant with expertise in COVID-19. Use the following texts and/or tables to produce a concise answer to the user question. \{user query\} \{docs\}.}}.
In cases where tables are included as part of the input, they are formatted to preserve structural integrity, accompanied by their respective captions and relevant in-text references to ensure that numerical or categorical data is properly contextualized.
For the interleaved configuration, the top 5 of each modality's retrieval ranking texts and tables are alternated in ranking order.

We aim to assess how well the language model integrates and synthesizes information from different retrieval modalities and whether the inclusion of tables impacts the quality and informativeness of the generated responses.


\subsection{Experimental Results}\label{sec:rag_results}

We employ a diverse set of metrics to evaluate both the retrieval performance of the configuration and the quality of the generated answer $as(q)$ for a given query $q$. Additionally, we conducted pairwise comparisons of answers to assess their usefulness, leveraging an Elo rating system for evaluation.

\paragraph{RAGAS-based Measures}
We evaluate the retrieval component using precision and recall derived from the synthetic relevance labels generated by REANIMATOR.
The labels were translated from four levels of relevance to binary relevance. Since the UMBRELA framework assigns relevance judgments on a scale from 0 to 3, we applied the same conversion method to ensure consistency when calculating binary inter-rater agreements and retrieval metrics that require binary relevance values. Specifically, relevance levels 0 and 1 were mapped to 0, while levels 2 and 3 were mapped to 1 \cite{umbrela}.
Additionally, we measure \emph{context relevance}, a RAGAS metric proposed by \citet{es_ragas_2023}, which gauges if the retrieved context $c(q)$ contains information that is needed to answer the question.
In particular, this metric aims to penalize the inclusion of redundant information.
The assessment uses \textit{gpt-4o-mini-2024-07-18}, prompting the model to extract a subset of sentences, $S_{ext}$, from $c(q)$ that are crucial to answer q.
The context relevance score is then computed as: $CR = \frac{\text{number of extracted sentences}}{\text{total number of sentences in c(q)}}$.

Faithfulness and answer relevance, two other RAGAS metrics, are used to evaluate the quality of the generated RAG output.
Faithfulness measures how well the generated answer aligns with the retrieved documents and tables~\cite{es_ragas_2023}: answer $as(q)$ is faithful to the context $c(q)$ if the claims that are made in the answer can be inferred from the context.
To estimate faithfulness, we first use \textit{gpt-4o-mini-2024-07-18} to extract a set of statements, $S(as(q))$.
The final faithfulness score, $F$, is then computed as $F = \frac{|V|}{|S|} $, where $|V|$ is the number of statements that were supported according to the LLM and $|S|$ is the total number of statements.

Answer relevance quantifies how well the generated answer addresses the input query~\cite{es_ragas_2023}.
Given an answer $as(q)$, we prompt an LLM to produce $n$ potential questions 
$\{q_i\}_{i=1}^{n}$ based on that answer.
We then embed both $q$ and each $q_i$ via the \texttt{text-embedding-3-small} model and compute $\mathrm{sim}(q, q_i)$ as the cosine similarity between their embeddings.
The final answer relevance score is:
\[
AR = \frac{1}{n}\sum_{i=1}^n \mathrm{sim}(q, q_i),
\]
which reflects the degree to which $as(q)$ addresses the query. 

\begin{figure}[t]

\subfloat[Retrieval Evaluation Metrics]{%
  \includegraphics[clip,width=\columnwidth]{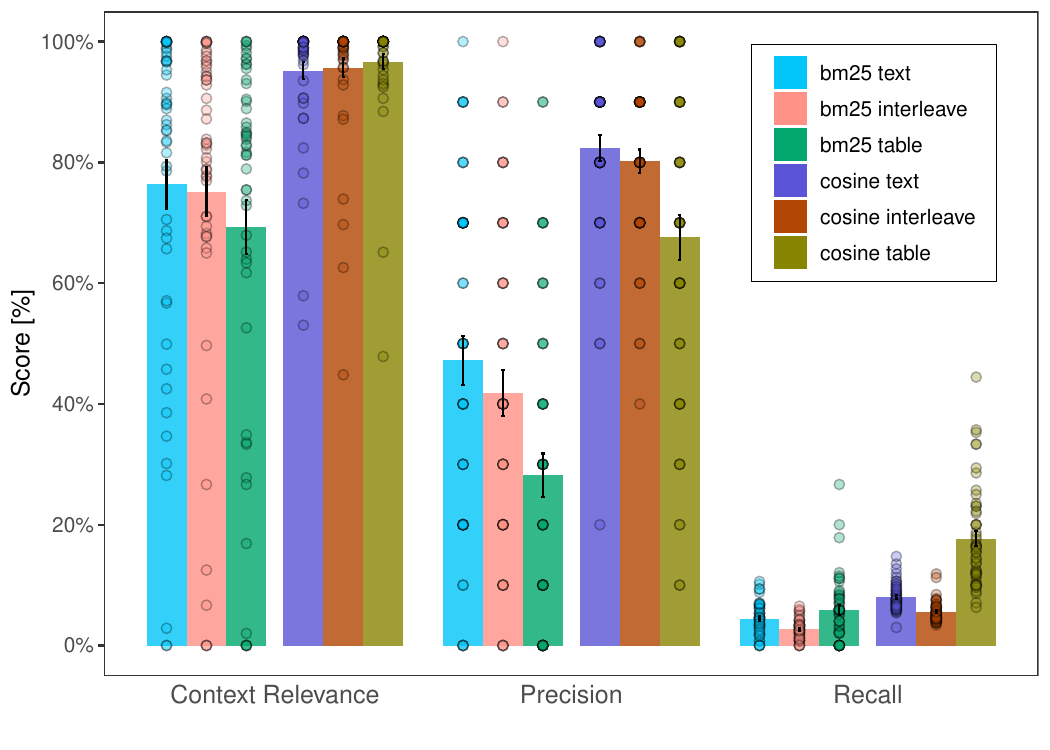}%
}

\subfloat[Generation Evaluation Metrics]{%
  \includegraphics[clip,width=1\columnwidth]{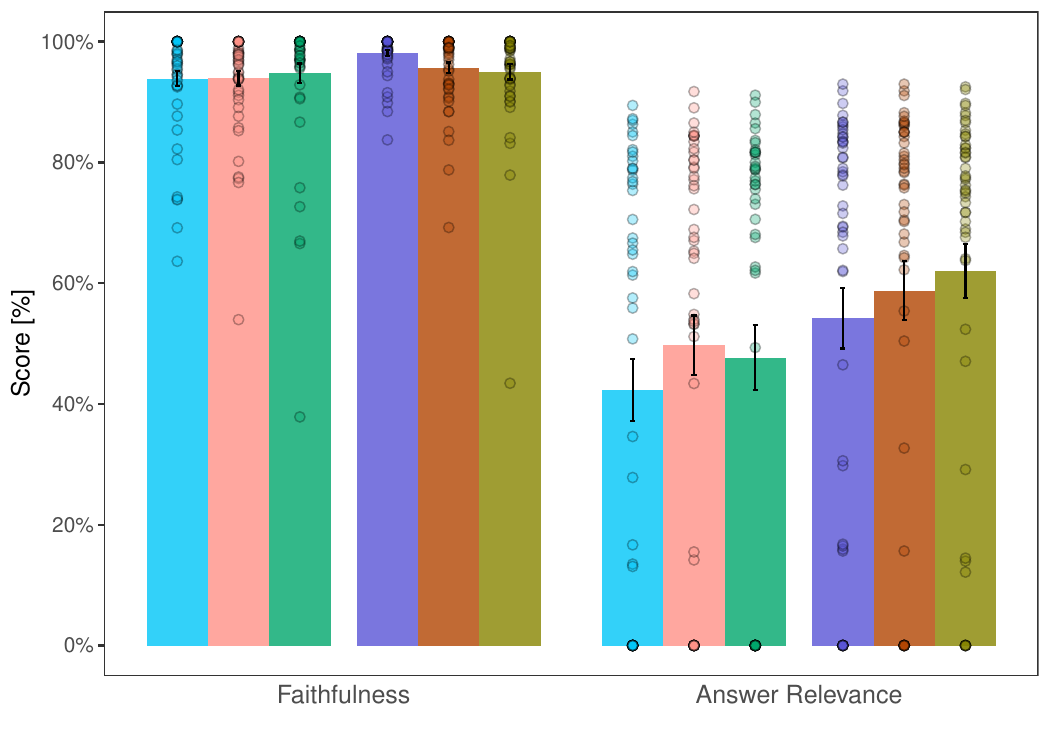}%
}

\caption{RAGAS and retrieval evaluation metrics of RAG with texts and tables.} 
\label{fig:results_RAGAS-metrics}
\end{figure}

\autoref{fig:results_RAGAS-metrics} present the results of the analysis of RAGAS and retrieval evaluation of RAG with texts and tables.
Cosine similarity retrieval outperforms BM25 retrieval notably, while the difference between the output of text-only, table-only, and combined RAG input is negligible.
This indicates that the LLM that generates the RAG output is as capable of generating text from tables as from texts.
It is noteworthy that answer relevance is generally lowered by many outputs deemed not relevant, occurring in all configurations.  

To better understand the impact of different retrieval modalities on the generated responses, we analyze the mean token count of the RAG outputs across the six configurations, reported in \autoref{tab:token_counts}.

\begin{table}[t]
    \centering
    \caption{Token Counts per Retrieval Configuration.}
    \label{tab:token_counts}
    \small
    \begin{tabular}{lc}
        \toprule
        \textbf{Retrieval Configuration} & \textbf{Token Count (Mean $\pm$ Std.)} \\
        \midrule
        BM25\textsubscript{text} & 226.5 $\pm$ 45.4 \\
        BM25\textsubscript{interleave} & 249.6 $\pm$ 57.1 \\
        Cosine\textsubscript{text} & 252.1 $\pm$ 31.3 \\
        BM25\textsubscript{table} & 274.9 $\pm$ 68.0 \\
        Cosine\textsubscript{interleave} & 283.5 $\pm$ 38.4 \\
        Cosine\textsubscript{table} & 322.9 $\pm$ 46.2 \\
        \bottomrule
    \end{tabular}
\end{table}

Results indicate that outputs generated from table-based retrieval configurations tend to be longer, with BM25\textsubscript{table} 
and Cosine\textsubscript{table} 
producing the longest responses, compared to text-only retrieval. 
Interleaved retrieval, which incorporates both text and tables, yields intermediate token counts. 
This increase in response length is likely due to the inherently higher character count of tables and their associated context (captions and in-text references), providing a richer and more structured information source for the language model.
While longer responses do not necessarily correlate with improved answer relevance or faithfulness, these results suggest that tables contribute additional content that the model incorporates into its outputs, potentially leading to greater detail or explanatory depth, reflecting in the generation evaluation metric scores.

A significant limitation of using RAGAS metrics to evaluate the impact of incorporating tables in RAG is that these metrics are based solely on the retrieved documents. Consequently, they do not assess the actual usefulness of the generated output for the user. Moreover, comparing systems that rely on different indices is inherently problematic, as the retrieved context originates from distinct information bases, making direct comparisons unreliable.

\paragraph{Pairwise Comparison and Elo-Based Ranking.}  
Assessing the overall usefulness of generated RAG outputs is complex and inherently subjective, as it depends not only on completeness, informativeness, and coherence but also on the recipient and the interpretation of the formulated query.  
Rather than attempting to assign absolute usefulness scores, we employ a pairwise comparison approach, enabling relative judgments between outputs.  
By systematically comparing pairs and aggregating results using an Elo algorithm~\cite{good_Elo}, we approximate a ranking that stabilizes over multiple iterations.  
Pairwise comparison is well suited to this task because direct usefulness judgments can be ambiguous, whereas relative judgments between two outputs tend to be more consistent \cite{gu2025surveyllmasajudge}.  
The Elo algorithm, originally developed for ranking chess players, assumes that a stronger player is more likely to win but allows for occasional upsets~\cite{boubdir2023elo}.  
Applied to RAG outputs, responses deemed more useful in repeated comparisons increase in rating, while less useful responses decrease. In recent years, this type of evaluation has become increasingly prevalent for assessing synthetic language generation \cite{chiang2024chatbotarenaopenplatform, köpf2023openassistantconversationsdemocratizing, bai2022traininghelpfulharmlessassistant, cui2024ultrafeedbackboostinglanguagemodels, engelmann-etal-2024-arts, 10.1145/3630744.3658415, gu2025surveyllmasajudge}.

We collect $5$ answers from each of the $6$ configurations across $50$ topics, yielding a total of $6\times5\times50=1500$ answers. For a given topic, with $30$ answers, the total number of pairwise comparisons is $\binom{30}{2}=435$. Excluding intra-configuration comparisons (with $\binom{5}{2}=10$ per configuration, hence $6\times10=60$), the valid comparisons per topic are $\binom{30}{2}-6\binom{5}{2}=375$, and over $50$ topics, this amounts to $375\times50=18750$ comparisons. 
Following the recommendations of Boubdir et al.~\cite{boubdir2023elo}, we enhance the reliability of our Elo scores by mitigating the match order dependency, which can otherwise lead to unreliable scores. To this end, we set the update parameter to $k=8$ which controls the magnitude of adjustments and compute the mean Elo score over 100 match-pair permutations. We implemented the pairwise comparison procedure using LangChain. 
Based on established recommendations \cite{gao_retrieval-augmented_2024, xiong_benchmarking_2024, yu_evaluation_2024} for evaluating the quality of RAG outputs, we selected the following criteria for pairwise preference assessment: Conciseness, Correctness, Coherence, Helpfulness, Depth, and Detail.

Initially, all outputs receive an initial Elo rating of 1500.  
For two responses, $T_1$ and $T_2$, with ratings $R_1$ and $R_2$, the expected probability of $T_1$ being judged superior is \cite{boubdir2023elo}:  

\begin{equation}
E_{T_1} = \frac{1}{1 + 10^{(R_2 - R_1)/400}}.
\end{equation}  

After the match outcome is determined and corresponding to the expected probability, $T_1$'s rating is updated as follows:

\begin{equation}
R_{1}' = R_{1} + k \bigl(S_{T_1} - E_{T_1}\bigr),
\end{equation}  

where $S_{T_1}$ is 1 if $T_1$ prevails and 0 otherwise.  
$T_2$’s rating is updated accordingly.  
After accumulating all pairwise judgments, the Elo scores reflect a stable ranking of output usefulness.  

\begin{figure}[t]
    \centering
    \includegraphics[width=1\linewidth]{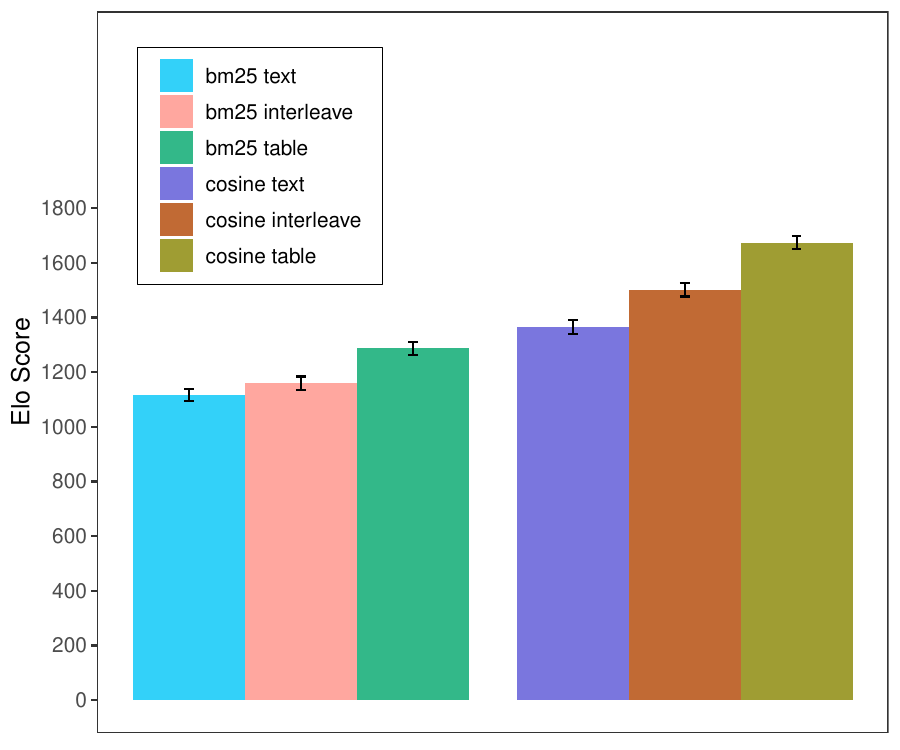}
    \caption{Elo scores for different retrieval configurations, ranking the usefulness of generated RAG outputs.} 
    \label{fig:elo_scores}
\end{figure}

The final Elo scores (see Figure~\ref{fig:elo_scores}) indicate that retrieval configurations incorporating tables produce more useful RAG outputs.  
Cosine\textsubscript{table} achieves the highest Elo score (1604.8), followed by Cosine\textsubscript{interleave} (1576.6), suggesting that structured table data enhances generated responses.  
Text-only retrieval ranks lower, with BM25\textsubscript{text} receiving the lowest Elo score (1189.9).  
BM25-based configurations also underperform their cosine similarity counterparts, reinforcing that embedding-based retrieval provides more useful context for generation.  

These findings align with our token count analysis (Section~\ref{sec:rag_setup}). 
While response length alone does not imply higher usefulness, incorporating tables into retrieval improves informativeness, particularly for applications requiring structured data, such as biomedical literature search.

\section{Conclusion and Outlook}

We address the challenge of re-using and expanding existing test collections by presenting REANIMATOR, a novel, flexible framework for automatically extracting resources like tables, captions, or in-text references from unstructured PDF documents and assigning synthetic relevance labels.
REANIMATOR can extend and revitalize existing test collections, allowing for the application of the test collection for different retrieval tasks (document retrieval, passage retrieval, table retrieval, RAG, and more) and application scenarios.

We showcase REANIMATOR's utility by revitalizing the TREC-COVID test collection into the augmented collection TREC-COVID+, making it suitable not only for document and passage retrieval but also table retrieval and retrieval-augmented generation (RAG).
We further show how such an enriched corpus can aid academic search tasks and how tables relevant to a given information need can be retrieved from an extensive literature corpus. 

Ultimately, REANIMATOR reduces the barriers to enriching existing test collections for new IR tasks. In particular, our Elo-based evaluation demonstrated that incorporating additional modalities for RAG tasks can be beneficial.
This repurposing aligns with the Green IR vision and the FAIR data principles, reducing the need for new, large-scale datasets and alleviating the computational overhead of training additional models from scratch.

These capabilities make REANIMATOR a versatile and accessible resource for the IR community, providing documented workflows and openly available resources that facilitate adoption and reproducibility. To the best of our knowledge, this is the first framework that automatically expands test collections with different modalities, enabling their application to multiple retrieval tasks such as table retrieval and retrieval-augmented generation. As interest in RAG and multi-modal retrieval grows, REANIMATOR is well-positioned to support a broad and expanding research community.


While our study provides valuable insights, certain aspects leave room for further refinement. Our implementation prioritized strong default settings rather than fine-tuning individual modules. Although this ensures robustness, more advanced techniques, such as improved semantic chunking and a more diverse pooling strategy, could further enhance the analysis of retrieval-based tasks. Although we incorporated human annotations, a more extensive annotation effort, particularly for pairwise preference comparisons, would improve reliability across different evaluation levels and provide deeper insights into system performance. In addition, our evaluation was performed within a specific use case, which may limit the generalizability of our findings. In particular, our approach does not fully capture the challenges associated with handling complex content, such as figures and equations. Finally, future work could explore automatic topic generation to improve scalability and refine the evaluation process, allowing for broad applicability across different domains.

\bibliographystyle{ACM-Reference-Format}
\bibliography{references}

\end{document}